\documentclass[conference]{IEEEtran}
\IEEEoverridecommandlockouts
\usepackage{cite}
\usepackage{amsmath,amssymb,amsfonts}
\usepackage{algorithmic}
\usepackage{graphicx}
\usepackage{textcomp}
\usepackage{xcolor}
\usepackage{tikz}
\usepackage{enumitem}
\usetikzlibrary{shapes.geometric, arrows}
\usepackage{algorithm}
\usepackage{hyperref}
\usepackage{booktabs}

\tikzstyle{startstop} = [rectangle, rounded corners, minimum width=3cm, minimum height=1cm,text centered, draw=black, fill=red!30]
\tikzstyle{process} = [rectangle, minimum width=3cm, minimum height=1cm, text centered, draw=black, fill=orange!30]
\tikzstyle{arrow} = [thick,->,>=stealth]

\def\BibTeX{{\rm B\kern-.05em{\sc i\kern-.025em b}\kern-.08em
    T\kern-.1667em\lower.7ex\hbox{E}\kern-.125emX}}
\begin{document}

\title{RoboSignature: Robust Signature and Watermarking on Network Attacks}

\author{
\IEEEauthorblockN{Aryaman Shaan\textsuperscript{*}}
\IEEEauthorblockA{\textit{New York University}\\
New York, USA \\
as12046@nyu.edu}
\and
\IEEEauthorblockN{Garvit Banga\textsuperscript{*} }
\IEEEauthorblockA{\textit{New York University}\\
New York, USA \\
gb2762@nyu.edu}
\and
\IEEEauthorblockN{Raghav Mantri\textsuperscript{*}}
\IEEEauthorblockA{\textit{New York University}\\
New York, USA \\
raghav.mantri@nyu.edu}
}

\maketitle
\begingroup\renewcommand\thefootnote{*}
\footnotetext{Equal contribution}
\endgroup
\begin{abstract}
Generative models have enabled easy creation and generation of images of all kinds given a single prompt. However, this has also raised ethical concerns about what is an actual piece of content created by humans or cameras compared to model-generated content like images or videos.

Watermarking data generated by modern generative models is a popular method to provide information on the source of the content.
The goal is for all generated images to conceal an invisible watermark, allowing for future detection and/or identification. The Stable Signature \cite{fernandez2023stablesignaturerootingwatermarks} finetunes the decoder of Latent Diffusion Models (LDM) such that a unique watermark is rooted in any image produced by the decoder.

In this paper, we present a novel adversarial fine-tuning attack that disrupts the model’s ability to embed the intended watermark, exposing a significant vulnerability in existing watermarking methods.

To address this, we further propose a tamper-resistant fine-tuning algorithm inspired by methods developed for large language models \cite{tamirisa2024}, tailored to the specific requirements of watermarking in LDMs. Our findings emphasize the importance of anticipating and defending against potential vulnerabilities in generative systems. Code for our proposed method is available at \url{https://github.com/GarvitBanga/RoboSignature}.

\end{abstract}

\section{Introduction}

Many recent generative models have the ability to watermark their content, and quite a few are open-source. However, individual developers can remove watermarks on generated images by simply removing a line of code in the inference section of the model \cite{fernandez2023stablesignaturerootingwatermarks}. This poses a challenge for companies and researchers that create the models and would like to make their models open source while ensuring that the models generate watermarks as part of their generation process. 

In an ideal scenario, the watermark should be inherently rooted in the model generation process, and the watermarks in the images/texts generated should be tamper-proof against image transformation attacks. Additionally, such open-source models themselves should be tamper-resistant, such that tampering with model weights does not allow generation of data without watermarks or incorrect watermarks. 

Fernandez et al. \cite{fernandez2023stablesignaturerootingwatermarks} explored various image-related generation tasks on LDMs and applied their novel fine-tuning method that embeds an invisible watermark in images during the LDM decoder's image generation step. These tasks include text-to-image, image-editing, and in-painting. However, they pointed out that the fine-tuned decoder could be subject to network-level attacks. The models could be subject to certain adversarial fine-tuning regime that would make the model `forget' to root the watermarks or root the incorrect watermark in the generation process while not harming the model's ability to generate images of good quality. The network levels attacks they pointed out are (i) model collusion and (ii) model purification. We introduce novel attacks based on random key generation that \textbf{confuses} the model such that it roots incorrect, random watermarks instead of the intended one. 

While contributions in attacking open-source models are important for exposing vulnerabilities to the wider research community to spread awareness of appropriate use cases, efforts are also needed to make models robust to such attacks. Tamirisa et al. \cite{tamirisa2024} observed that safeguards enabled in Large Language Models (LLMs) can be easily bypassed using tampering attacks. These attacks typically involve a few steps of fine-tuning and have been proven effective in circumventing safeguards designed to prevent harmful information disclosure and unlearning. To mitigate such a threat, they proposed a novel fine-tuning algorithm to make LLMs tamper-resistant to adversarial fine-tuning while retaining their ability to answer informative questions and perform day-to-day tasks.

A significant aspect of our contribution involves adapting the approach proposed by Tamirisa et al. \cite{tamirisa2024} for LLMs to LDMs to enhance the watermark rooting procedure developed by Fernandez et al. \cite{fernandez2023stablesignaturerootingwatermarks}, making it robust to the adversarial fine-tuning attacks we introduce. Importantly, our approach ensures that the core image generation capabilities of the model remain intact, preserving its utility for practical applications.

\section{Related Work}

\subsection{Image Generation}
Diffusion models are a class of generative models capable of producing high-quality images by learning the data distribution through a process of noise addition and removal \cite{DBLP:journals/corr/abs-2006-11239}. During the forward process (encoding), noise is progressively added to the input data over multiple steps, gradually transforming it into a noise-like representation. In the reverse process (decoding), a trained model iteratively removes the noise in small steps, reconstructing data samples that closely resemble the original distribution. This iterative de-noising process allows diffusion models to generate images with remarkable detail and fidelity.

Latent Diffusion Models (LDMs) extend this approach by operating in a compressed latent space rather than directly on high-dimensional image data \cite{Rombach_2022_CVPR}. An auto-encoder structure is used to map images into a lower-dimensional latent representation, where the diffusion process takes place. Noise is added to these latent representations during the forward process, and the decoder iteratively removes this noise to reconstruct the data. Once trained, only the latent space and decoder are required for image generation, enabling LDMs to achieve efficient and high-quality results.

StableDiffusion is a specific LDM-based model\cite{Rombach_2022_CVPR}. that we use for our research. It is capable of generating images based on textual prompts. 

\subsection{Watermarking}

\subsubsection{HiDDeN}
A classical work in deep learning image watermarking literature is Hiding Data with Deep Networks (HiDDeN) \cite{zhu2018hiddenhidingdatadeep}. Two sets of parameters are jointly optimized, the encoder network $W_E$ and the extractor network $W$. $W_E$ embeds a k-bit message into images robustly, while various adversarial transformation are applied to images during images. $W_E$ is trained to extract the original k-bit message. This work was inspired from the fact that minor noise perturbations can be added to images, that can be detected by neural networks, but are impervious to the naked eye. 

Formally $W_E$ takes as input an image $i_o$ and a watermark $ m \in \{0,1\}^k$. It outputs a residual image $\beta$ (essentially a minor noise perturbation) that is the same size as $i_o$. The residual image is scaled by a factor of $\alpha$ to produce an image with a watermark $i_w = i_o + \alpha \beta$. Next, a transformation is applied to $i_w$ from a set of transformations $T$. Examples are cropping and JPEG compression. $W$ then extracts a soft message from the transformed image, $m' = W(T(i_w))$. To train $W$ effectively, binary cross-entropy (BCE) loss is used between the k-bit message $m$ and the sigmoid $\sigma(m')$. The formula for BCE loss is -
$$L_m = -\sum_{i=1}^{k} m_i \cdot \log(\sigma(m'_i)) + (1 - m_i) \cdot \log(1 - \sigma(m'_i))$$.

\subsubsection{Stable Signature}

Fernandez et al. \cite{fernandez2023stablesignaturerootingwatermarks} proposed Stable Signature, an active strategy combining image watermarking and Latent Diffusion Models. First, a HiDDen network is trained on the COCO dataset \cite{DBLP:journals/corr/LinMBHPRDZ14}. The encoder of this network is discarded. Only the extractor is used. Additionally, a particular k-bit string is chosen beforehand. This acts as the watermark key.

Next, a pre-trained LDM model is chosen and the decoder of the model is fine-tuned. Specificaly, the encoder and the latent space layer weights are frozen during the backpropogation step. The decoder of a stable diffusion model is now fine-tuned to generate images with the watermarks encoded. Images generated by the LDM decoder are then fed into the HiDDen watermark extractor and an encoded message/watermark is extracted from them. This extracted watermark must match the watermark that was chosen before. The decoder essentially needs to be fine-tuned to embed the original watermark into any image it generates. This is ensured by the loss function.

This process will be fine-tuned with 2 key components in the loss function given by

$$L = L_m + L_i$$

where $L_m$ is the loss generated from the difference between the original watermark signature and the extracted watermark, and $L_i$ is the loss component derived from the difference between the generated image from this fine-tuned decoder $D_m$ and the original decoder with no fine-tuning $D$. This is to ensure that the image generation with the message is not radically different from images originally generated images without watermarks.

It is also crucial that the model is able to understand luminance and contrast-masking to produce less perceivable watermarks. To do this, the decoder of the original Stable Diffusion model is used to produce an image without a watermark. The image perceptual loss between this image and the image produced with watermark controls the distortion between the images. This image perceptual loss is added along with the loss generated from the difference between the original watermark signature and the extracted watermark to train the weights of the decoder $D_m$.

Any image produced by the LDM decoder would have that specific k-bit message embedded in it. This k-bit message could possibly be unique to that LDM decoder, as such an image produced could be traced back to the user in possession of that particular LDM decoder. 

\subsection{Network Level Attacks}

Fernandez et al. \cite{fernandez2023stablesignaturerootingwatermarks} tested for various image level attacks that tried to remove the watermark from generated images will trying to keep image quality intact. They also pointed out an important class of attacks - network level attacks. Where the LDM decoder's weights could be modified in a manner such that the model either forgets to root watermarks in generated image, or roots the incorrect watermark. The attacker's objective, in addition to corrupting the watermarking capability of the model, also includes preserving the quality of the generated images. There are primary two network level attacks discussed by Fernandez et al. \cite{fernandez2023stablesignaturerootingwatermarks} Such attacks are primarily adversarial fine-tuning based methods. 

\textbf{The model collusion} attack is caused when multiple users that posses different LDM decoders collude. They can average the weights of their models to create a new model that generates images of a high quality but produces a new unique watermark key. This would help them escape the identification process and images produced would not trace back to them. 

\textbf{The model purification} attack aimed at making the LDM decoder forget to add the watermark in its generation process. Essentially the model is fine-tined again but without the loss pertaining to the message-key. The idea is for the model to leave out the key generation. Fernandez et al. \cite{fernandez2023stablesignaturerootingwatermarks} experimentally showed that such an attack starts reducing the bit accuracy of the watermark extracted after a significant number of training steps. Also while the the bit accuracy reduces the peak signal to noise ratio (PSNR) value also reduces. Hence, image quality does not remain intact. 

\subsection{Tamper Resistant Fine-Tuning in LLMs (TAR)}

Tamirisa et al. \cite{tamirisa2024} made the observation that a few steps of adversarial fine-tuning could remove safety guardrails set in LLMs. They mainly focused on guardrails around Weaponization knowledge restriction and harmful request refusal. Weaponization knowledge restriction refers to a guardrail that ensures that the language model does not relay information regarding the use and manufacture of weapons while answering prompts that only request information for benign domains. Similarly, the guardrail of harmful request refusal disallows the model to produce harmful outputs. 
Tamirisa et al. \cite{tamirisa2024} developed a novel fine-tuning algorithm called tamper resistant fine-tuning (TAR) that ensures any future adversarial fine-tuning does not succeed in removing guardrails already set in the model. They define the objective of TAR to keep the safety metric and the capability metric of the model high post fine-tuning. The capability metric of the model is defined as the performance of the model in generation, for example producing outputs pertaining to benign knowledge domains. The safety metric refers to the model's ability to adhere to the given guardrail. Formally, let the vanilla model's weights weights be $\theta$, and let the guardrails be $G$. After guardrails have been set in the model, its weights are then transformed to $\theta_G$. Post this TAR fine-tuning is performed on the model to produce $\theta_G'$. The aim is that after attacks from a set $A_{test}$ are applied to $\theta_G'$, $\mathrm{capabilities\_metric}(\theta_G')$ and $\mathrm{safety\_metric}(\theta_G')$ remain high.

\section{Methods}

\subsection{Proposed Attacks}
We propose a novel adversarial fine-tuning attack on StableDiffusion's decoder which has been previously fine-tuned by the Stable Signature method to root watermarks. 

The \textbf{Random Key Attack} is a fine-tuning regime where in each training step, for $L_m$, the loss pertaining to message bit accuracy, the ground truth $k$-bit message is chosen randomly from the key space. In our work, and in Ferenandez et al.'s\cite{fernandez2023stablesignaturerootingwatermarks} approach, the watermark key is a bit string of size $48$, leading to a key space size is $2^{48}$. This fine-tuning regime is intended to confuse the model, causing it to generate images embedded with random keys rather than the intended watermark key. 

The \textbf{Gradual Random Key Attack} is a variant of the Random Key Attack. Instead of producing a completely random key at every training step, the key is gradually modified, starting with a single random bit change and progressing towards a fully random key. The intent remains the same: the model should generate images with an incorrect embedded watermark key as the attack progresses.

\subsection{Modified TAR Fine-Tuning}
To make the model robust against adversarial fine-tuning, we adapt the Tampering Attack Resistance (TAR) proposed by Tamirisa et al. \cite{tamirisa2024} to fit our framework. 

We split the training data into two datasets: the \textbf{ capabilities\_metric proxy dataset} $\mathcal{D}_{\text{retain}}$ and the \textbf{safety\_metric proxy dataset} $\mathcal{D}_{\text{TR}}$. Theses datasets are used to train two separate sets of gradients,  $g_{TR}$ and $g_{retain}$. During each outer training step, both sets of weights are updated. 

The $g_{TR}$ gradients are computed by attacking the current iteration of the model, ($\theta_{i-1}$), using $K$ Random Key Attacks. Each attack consists of approximately $100$ training steps. The attacked model obtained through these steps is denoted as $\text{attack}(\theta_{i-1})$. For each attacked model, we calculate the loss between the model’s output and the original target key ($L_m$). Additionally, we compute the image reconstruction loss ($L_i$), which measures the difference between the images generated by the attacked model and those generated by the original model after Stable Signature fine-tuning. These losses are combined to form the total loss $L_{TR}$ (as described in Algorithm $1$). The $g_{TR}$ gradients are then computed as the average of the gradients from each $L_{TR}$ loss.

Similarly, the $g_{retain}$ gradients are computed using the original loss function, $L_{ORG}$. This loss ensures the output key matches the original target key (without any random attacks). The image reconstruction loss is also included as a component of $L_{ORG}$. Additionally, the L2 norm of the difference between the hidden states of the current iteration of the model and the original model is incorporated into the $g_{retain}$ gradients.

\begin{algorithm}[H]
\caption{TAR: Tampering Attack Resistance}
\begin{algorithmic}[1]
\REQUIRE Initial LDM Decoder parameters $\theta$,  capabilities\_metric proxy dataset $\mathcal{D}_{\text{retain}}$, safety\_metric proxy dataset $\mathcal{D}_{\text{TR}}$, outer steps $N$, learning rate $\eta$, $h_\theta(\cdot)$ returns the residual stream hidden states for model parameters $\theta$
\STATE $\theta_0 \leftarrow$ Apply Initial Stable Signature fine-tuning to $\theta$
\FOR{$i = 1$ to $N$}
    \STATE $g_{\text{TR}} \leftarrow 0$ \hfill \# For accumulating tamper-resistance gradient
    \STATE Sample $x_{\text{TR}} \sim \mathcal{D}_{\text{TR}}$
    \FOR{$k = 1$ to $K$}
        \STATE Sample random key for $\text{attack}$
        \STATE $g_{\text{TR}}\leftarrow g_{\text{TR}} + \frac{1}{K} \nabla_{\theta_{i-1}} L_{\text{TR}}(\text{attack}(\theta_{i-1}), x_{\text{TR}})$  \# Tamper-resistance loss
    \ENDFOR
    \STATE Sample $x_r \sim \mathcal{D}_{\text{retain}}$
    \STATE $g_{\text{retain}} \leftarrow \nabla_{\theta_{i-1}} \left[ L_{\text{ORG}}(\theta_{i-1}, x_r) + \|h_{\theta_{i-1}}(x_r) - h_\theta(x_r)\|_2^2 \right]$ \hfill \# Retain loss
    \STATE $\theta_i \leftarrow \theta_{i-1} - \eta \left[ \lambda_{\text{TR}} \cdot g_{\text{TR}} + \lambda_{\text{retain}} \cdot g_{\text{retain}} \right]$  \# Update parameters
\ENDFOR
\STATE $\theta_G \leftarrow \theta_N$
\RETURN $\theta_G$
\end{algorithmic}
\end{algorithm}

At the end of each training step, the current iteration of the model ($\theta_{i-1}$) is updated with both $g_{TR}$ and $g_{retain}$ gradients. Essentially, $g_{TR}$ instructs the model on how to recover after being attacked, enhancing the safety metric of the model. Meanwhile, $g_{retain}$ reinforces the model’s ability to produce the target watermark key while maintaining high image quality.

\section{Experimental Results}
\subsection{Results of Proposed Attacks}
We conducted adversarial fine-tuning experiments on an LDM decoder that was previously fine-tuned to embed a specific watermark in the images it generates.
These attacks aim to generate - 
\begin{enumerate}
    \item Random keys at each step (Table ~\ref{tab:strategy1})
    \item Gradually random or different keys across steps (Table ~\ref{tab:strategy2})
\end{enumerate}
The fine-tuning process for the LDM decoder involved $100$ steps with a batch size of $4$, utilizing $400$ images from the COCO dataset, similar to the setup described by Fernandez et al. \cite{fernandez2023stablesignaturerootingwatermarks}.

\begin{table}[h!]
\centering
\caption{Random keys in each training step (Strategy 1)}
\label{tab:strategy1}
\begin{tabular}{@{}lcc@{}}
\toprule
\textbf{Type} & \textbf{PSNR} & \textbf{Bit Accuracy} \\ \midrule
Train         & 27.922440     & 0.503542              \\ 
Eval          & 27.599785     & 0.563099              \\ \bottomrule
\end{tabular}
\end{table}

\begin{table}[h!]
\centering
\caption{Gradual randomness (Strategy 2)}
\label{tab:strategy2}
\begin{tabular}{@{}lcc@{}}
\toprule
\textbf{Type} & \textbf{PSNR} & \textbf{Bit Accuracy} \\ \midrule
Train         & 27.947460     & 0.521354              \\ 
Eval          & 27.971604     & 0.537864              \\ \bottomrule
\end{tabular}
\end{table}
The low bit accuracies but high PSNR values highlight the dangers of such adversarial attacks - despite maintaining high image quality, as indicated by PSNR values, the bit accuracies dropped significantly which implies that a user would be unable to reliably trace the generated images back to the correct model of origin. This confirmed our hypothesis of the original model being susceptible to adversarial fine-tuning/network-level attacks. Notably, a $50\%$ bit accuracy corresponds to completely random keys being generated. Thus, bit accuracies close to $50\%$ indicate that the model is effectively incapable of rooting the intended target watermark key.

\subsection{Results of Modified TAR fine-tuning}

We conducted our modified TAR fine-tuning regime on LDM decoders that were already fine-tuned using the Stable Signature method for embedding watermarks. We used different configurations of outer training steps ($N$), different number of attacks to calculate $g_{TR}$ ($K$) and different number of train steps during each attack (attack steps) as shown in table \ref{tab:evaluation_results}. 

\begin{table}[h!]
\centering
\caption{Evaluation results for different Outer Steps ($N$), Inner Steps ($K$), and Attack Steps.}
\label{tab:evaluation_results}
\resizebox{\columnwidth}{!}{%
\begin{tabular}{@{}lccccc@{}}
\hline
\textbf{Outer Steps ($N$)} & \textbf{Inner Steps ($K$)} & \textbf{Attack Steps} & \textbf{PSNR (eval)} & \textbf{Bit Accuracy (eval)} \\ \hline
20                       & 20                       & 100                   & 51.949070            & 0.995949                    \\ 
50                       & 20                       & 50                    & 35.128624            & 0.963542                    \\ 
50                       & 20                       & 100                   & 42.638180            & 0.994027                    \\ 
\hline
\end{tabular}%
}
\end{table}

After our modified TAR fine-tuning, high PSNR values (ranging from $35 – 51$) were observed on the evaluation dataset, indicating that high-quality images were consistently generated. The bit accuracy relative to the target key was also notably high ($96 – 99\%$) on the evaluation dataset across various training configuration

\begin{figure}[h!]
    \centering
    \includegraphics[width=0.5\textwidth]{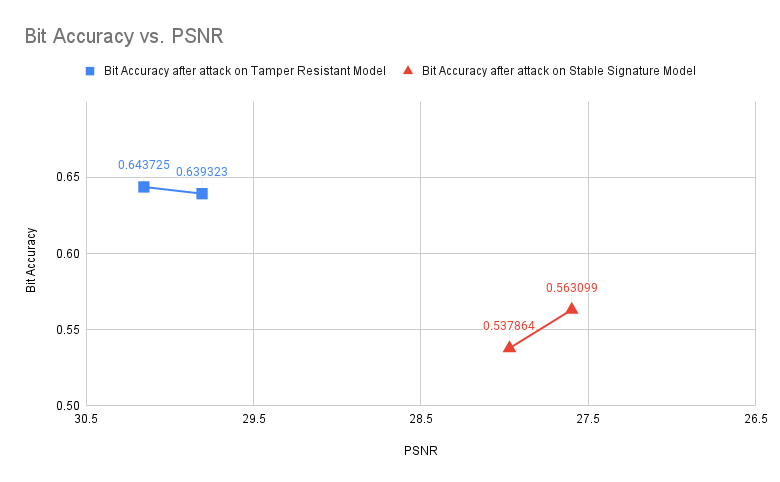}
    \caption{Bit Accuracy vs PSNR}
    \label{fig:bit_accuracy_vs_psnr}
\end{figure}

We further subjected the fine-tuned models to Random Key Attacks. Post-attack, the bit accuracy relative to the target key dropped to approximately $64–65\%$, while PSNR values remained in the range of $29–32$.These values are depicted using the square markers in Figure \ref{fig:bit_accuracy_vs_psnr}. For comparison, Figure \ref{fig:bit_accuracy_vs_psnr} also shows the bit accuracy and PSNR values for the original Stable Signature LDM decoder post-attack, shown using triangle markers. 

\section{Conclusion}
Our analysis highlights that achieving tamper resistance for watermarking in Latent Diffusion Models (LDMs) is a tractable yet highly intricate challenge. While our proposed tamper-resistant fine-tuning method demonstrates minor improvements in bit accuracy against various tampering attacks, these enhancements fall short of addressing all vulnerabilities, especially in scenarios involving complex image transformations.

A significant observation from our study is the absence of any underlying distribution among random bit strings.  The possible  key space for bit-strings is exceptionally large $(2^{48})$, with only one target key correct out of these possibilities. This makes the task of tamper resistance considerably challenging when compared to tasks like mitigating harmful content in the context of LLMs, where some inherent redundancy or distribution aids in the defense.

Further, unlike the robust safeguards established in the original TAR framework for LLMs, LDMs lack an equivalent baseline defense mechanism, leaving them inherently more susceptible to adversarial manipulations. Developing an effective initial safeguard for LDMs remains an open problem, underscoring the non-trivial nature of ensuring security and resilience in this domain.

\section{Acknowledgments}
We express our sincere gratitude to Prof. Saining Xie for his insightful teachings on Computer Vision, and giving us time to refine our research idea. We'd also like to thank our tutors of the course for their guidance, and NYU HPC for providing us the necessary compute to conduct our experiments.

\section{Contribution statement}

All team members have contributed equally to this project.

\bibliographystyle{IEEEtran}
\bibliography{biblography}

\begin{thebibliography}{1}
\providecommand{\url}[1]{#1}
\csname url@samestyle\endcsname
\providecommand{\newblock}{\relax}
\providecommand{\bibinfo}[2]{#2}
\providecommand{\BIBentrySTDinterwordspacing}{\spaceskip=0pt\relax}
\providecommand{\BIBentryALTinterwordstretchfactor}{4}
\providecommand{\BIBentryALTinterwordspacing}{\spaceskip=\fontdimen2\font plus
\BIBentryALTinterwordstretchfactor\fontdimen3\font minus \fontdimen4\font\relax}
\providecommand{\BIBforeignlanguage}[2]{{%
\expandafter\ifx\csname l@#1\endcsname\relax
\typeout{** WARNING: IEEEtran.bst: No hyphenation pattern has been}%
\typeout{** loaded for the language `#1'. Using the pattern for}%
\typeout{** the default language instead.}%
\else
\language=\csname l@#1\endcsname
\fi
#2}}
\providecommand{\BIBdecl}{\relax}
\BIBdecl

\bibitem{fernandez2023stablesignaturerootingwatermarks}
\BIBentryALTinterwordspacing
P.~Fernandez, G.~Couairon, H.~Jégou, M.~Douze, and T.~Furon, ``The stable signature: Rooting watermarks in latent diffusion models,'' 2023. [Online]. Available: \url{https://arxiv.org/abs/2303.15435}
\BIBentrySTDinterwordspacing

\bibitem{tamirisa2024}
\BIBentryALTinterwordspacing
R.~Tamirisa, B.~Bharathi, L.~Phan, A.~Zhou, A.~Gatti, T.~Suresh, M.~Lin, J.~Wang, R.~Wang, R.~Arel, A.~Zou, D.~Song, B.~Li, D.~Hendrycks, and M.~Mazeika, ``Tamper-resistant safeguards for open-weight llms,'' 2024. [Online]. Available: \url{https://arxiv.org/abs/2408.00761}
\BIBentrySTDinterwordspacing

\bibitem{DBLP:journals/corr/abs-2006-11239}
\BIBentryALTinterwordspacing
J.~Ho, A.~Jain, and P.~Abbeel, ``Denoising diffusion probabilistic models,'' \emph{CoRR}, vol. abs/2006.11239, 2020. [Online]. Available: \url{https://arxiv.org/abs/2006.11239}
\BIBentrySTDinterwordspacing

\bibitem{Rombach_2022_CVPR}
R.~Rombach, A.~Blattmann, D.~Lorenz, P.~Esser, and B.~Ommer, ``High-resolution image synthesis with latent diffusion models,'' in \emph{Proceedings of the IEEE/CVF Conference on Computer Vision and Pattern Recognition (CVPR)}, June 2022, pp. 10\,684--10\,695.

\bibitem{zhu2018hiddenhidingdatadeep}
\BIBentryALTinterwordspacing
J.~Zhu, R.~Kaplan, J.~Johnson, and L.~Fei-Fei, ``Hidden: Hiding data with deep networks,'' 2018. [Online]. Available: \url{https://arxiv.org/abs/1807.09937}
\BIBentrySTDinterwordspacing

\bibitem{DBLP:journals/corr/LinMBHPRDZ14}
\BIBentryALTinterwordspacing
T.~Lin, M.~Maire, S.~J. Belongie, L.~D. Bourdev, R.~B. Girshick, J.~Hays, P.~Perona, D.~Ramanan, P.~Doll{\'{a}}r, and C.~L. Zitnick, ``Microsoft {COCO:} common objects in context,'' \emph{CoRR}, vol. abs/1405.0312, 2014. [Online]. Available: \url{http://arxiv.org/abs/1405.0312}
\BIBentrySTDinterwordspacing

\end{thebibliography}

\end{document}